\documentclass[journal=jacsat,manuscript=article]{achemso}
\usepackage[version=3]{mhchem}
\usepackage{hyperref}
\usepackage{amsmath}
\usepackage{xcolor}
\usepackage{booktabs}
\usepackage{xr}
\usepackage{siunitx}

\newcolumntype{H}{>{\setbox0=\hbox\bgroup}c<{\egroup}@{}}

\newif\ifhighlightnumbers
\highlightnumbersfalse 
\newcommand{\numericalresult}[1]{\ifhighlightnumbers \textcolor{red}{#1}\else #1\fi}

\makeatletter
\newcommand*{\addFileDependency}[1]{
  \typeout{(#1)}
  \@addtofilelist{#1}
  \IfFileExists{#1}{}{\typeout{No file #1.}}
}
\makeatother

\newcommand*{\myexternaldocument}[1]{%
    \externaldocument{#1}%
    \addFileDependency{#1.tex}%
    \addFileDependency{#1.aux}%
}

\myexternaldocument{nmr-supplement}

\author{Frank Hu}
\author{Michael S. Chen}
\author{Grant M. Rotskoff}
\email{rotskoff@stanford.edu}
\author{Matthew W. Kanan}
\email{mkanan@stanford.edu}
\author{Thomas E. Markland}
\email{tmarkland@stanford.edu}
\affiliation[Stanford University]
{Department of Chemistry, Stanford University}

\title[NMR structure elucidation]
  {Accurate and efficient structure elucidation from routine one-dimensional NMR spectra using multitask machine learning}

\abbreviations{IR,NMR}

\begin{document}

\begin{abstract}
Rapid determination of molecular structures can greatly accelerate workflows across many chemical disciplines. However, elucidating structure using only one-dimensional (1D) NMR spectra, the most readily accessible data, remains an extremely challenging problem because of the combinatorial explosion of the number of possible molecules as the number of constituent atoms is increased. Here, we introduce a multitask machine learning framework that predicts the molecular structure (formula and connectivity) of an unknown compound solely based on its 1D \textsuperscript{1}H and/or \textsuperscript{13}C NMR spectra. First, we show how a transformer architecture can be constructed to efficiently solve the task, traditionally performed by chemists, of assembling large numbers of molecular fragments into molecular structures. Integrating this capability with a convolutional neural network (CNN), we build an end-to-end model for predicting structure from spectra that is fast and accurate. We demonstrate the effectiveness of this framework on molecules with up to 19 heavy (non-hydrogen) atoms, a size for which there are trillions of possible structures. Without relying on any prior chemical knowledge such as the molecular formula, we show that our approach predicts the exact molecule \numericalresult{69.6\%} of the time within the first 15 predictions, reducing the search space by up to 11 orders of magnitude. 
\end{abstract}

\section{Introduction}
 
1D \textsuperscript{1}H and \textsuperscript{13}C~NMR spectroscopy are workhorse methods in structure elucidation since they are both easy to perform and provide a wealth of information about molecular composition, connectivity, and stereochemistry. However, interpretation of NMR spectra remains time-intensive and error-prone, often requiring high levels of chemical expertise and prior knowledge that constrains the space of possible structures. For small molecules or structures that are composed of a known subset of chemical building blocks, forward prediction methods\cite{binev_prediction_2007,meiler_proshift_nodate,gao_general_2020,li_highly_2024} can be used to compare with databases of spectra, obtained either from previous experiments or simulations, with the hope of obtaining a match with the measured spectra\cite{wu_elucidating_2023,lemm_impact_2024,wei_deep_2022,sun_cross-modal_2024,smith_assigning_2010,elyashberg_computer_2021,elyashberg_enhancing_2023,wenk_sherlockfree_2023,howarth_dp4-ai_2020}. However, the number of possible molecules suffers from a combinatorial explosion as the number of constituent atoms increases. For example, for organic compounds made up of up to 11 heavy (non-hydrogen) atoms (C, N, O, S, and/or halogen), there are $\sim$26 million molecules consistent with the basic bonding rules of chemistry (not including stereoisomers), but by 17 heavy atoms this number grows to $\sim$166 billion possible structures \cite{ruddigkeit_enumeration_2012}, and by 21 heavy atoms exceeds 20 trillion. Addressing this formidable challenge to unlock unsupervised structure elucidation of organic molecules would remove a key bottleneck in chemical research and could be paired with automated synthetic workflows\cite{vriza_self-driving_2023,xie_toward_2023,szymanski_autonomous_2023} to create closed-loop discovery platforms.

We and others have recently reported machine learning (ML) approaches to structure elucidation based on identifying molecular substructures (functional groups, small fragments) from NMR spectral data and using this information to assemble a molecular structure\cite{li_identifying_2022,specht_automated_2024,huang_framework_2021} or based on the task of selecting which of a set of provided candidate molecules corresponds to a given NMR spectrum and assigning the peaks in that provided spectrum to specific atoms\cite{schilter_unveiling_2023}. While ML models have proven to be remarkably effective in identifying substructures, converting substructure information to a molecular structure is much more challenging. Beam searching over possible structures\cite{huang_framework_2021} or building the structure in an atom-by-atom manner\cite{sridharan_deep_2022,devata_deepspinn_2024} are viable for small systems but quickly lose effectiveness as the number of atoms increases because of the combinatorial scaling of the problem size.

Rather than predicting fragments as an intermediate step, recent work has sought to directly predict molecular structure from spectra in an end-to-end fashion, using deep learning architectures such as graph convolutional neural networks\cite{sapegin_structure_2024} and transformers\cite{yao_conditional_2023,alberts_learning_2023,alberts_leveraging_2023}. For molecules with up to 13 heavy atoms a top-10 accuracy of 78.5\% was achieved with a transformer model using as inputs the molecular formula and infrared (IR) spectra that were pre-processed into a sequence of integers representing the intensity\cite{alberts_leveraging_2023}. On systems of up to 35 heavy atoms, a similar approach with transformers yielded a top-10 accuracy of 86.6\% using the molecular formula and \textsuperscript{1}H~NMR and \textsuperscript{13}C~NMR spectra pre-processed into strings containing information about peak shifts, splittings, and multiplicities \cite{alberts_learning_2023}. For larger organic systems ranging from a few dozen to over a hundred heavy atoms, it was shown that one can obtain a top-10 accuracy of 94.2\% when feeding the molecular formula, \textsuperscript{13}C~NMR shifts, and a SMILES\cite{weininger_smiles_1988} string representation of a large fragment of the target molecule into a transformer pre-trained on 360 million molecules\cite{yao_conditional_2023}. These studies suggest that end-to-end frameworks are a promising approach to address the combinatorial problem when extensive pre-processing is applied to the spectra and sufficient chemical information is provided (e.g., the molecular formula). However, in many cases information such as the molecular formula is not readily available, and pre-processing spectra is burdensome and introduces biases. It is therefore critical to develop a framework that can accurately and efficiently predict the molecular structure of an unknown compound using raw spectral data alone. 

Here, we present a transformer-based ML framework for solving the most challenging version of the structure elucidation problem, applying minimal pre-processing to the input \textsuperscript{1}H and \textsuperscript{13}C NMR spectra and using no other prior information such as molecular formula or molecular fragments. First, we train a transformer model\cite{vaswani_attention_2017} to solve the problem of constructing the molecular structure (formula and connectivity) when given only information about the presence or absence of a set of 957 very simple substructures ($\leq$7 atoms). We show that a transformer architecture recovers the exact molecular structure with high accuracy, succeeding \numericalresult{93.2\%} of the time within the first 15 predictions when tested on molecules with up to 19 heavy atoms. Next, we integrate this pre-trained transformer into a multitask model that predicts both substructure and molecular structure. This end-to-end model inputs only 1D \textsuperscript{1}H~NMR and \textsuperscript{13}C~NMR spectra and yields the correct molecular structure \numericalresult{69.6\%} of the time within the first 15 predictions when tested on simulated spectra for molecules with up to 19 heavy atoms. This model is thus capable of rapidly constraining the massive chemical search space ($>$2 trillion possibilities) using just routinely available NMR spectra, providing a complementary tool to other existing structure elucidation, reaction prediction, and retrosynthesis frameworks. 

\section{Results and discussion}\label{sec:results_discussion}
\begin{figure}
    \centering
    \includegraphics[width=\textwidth]{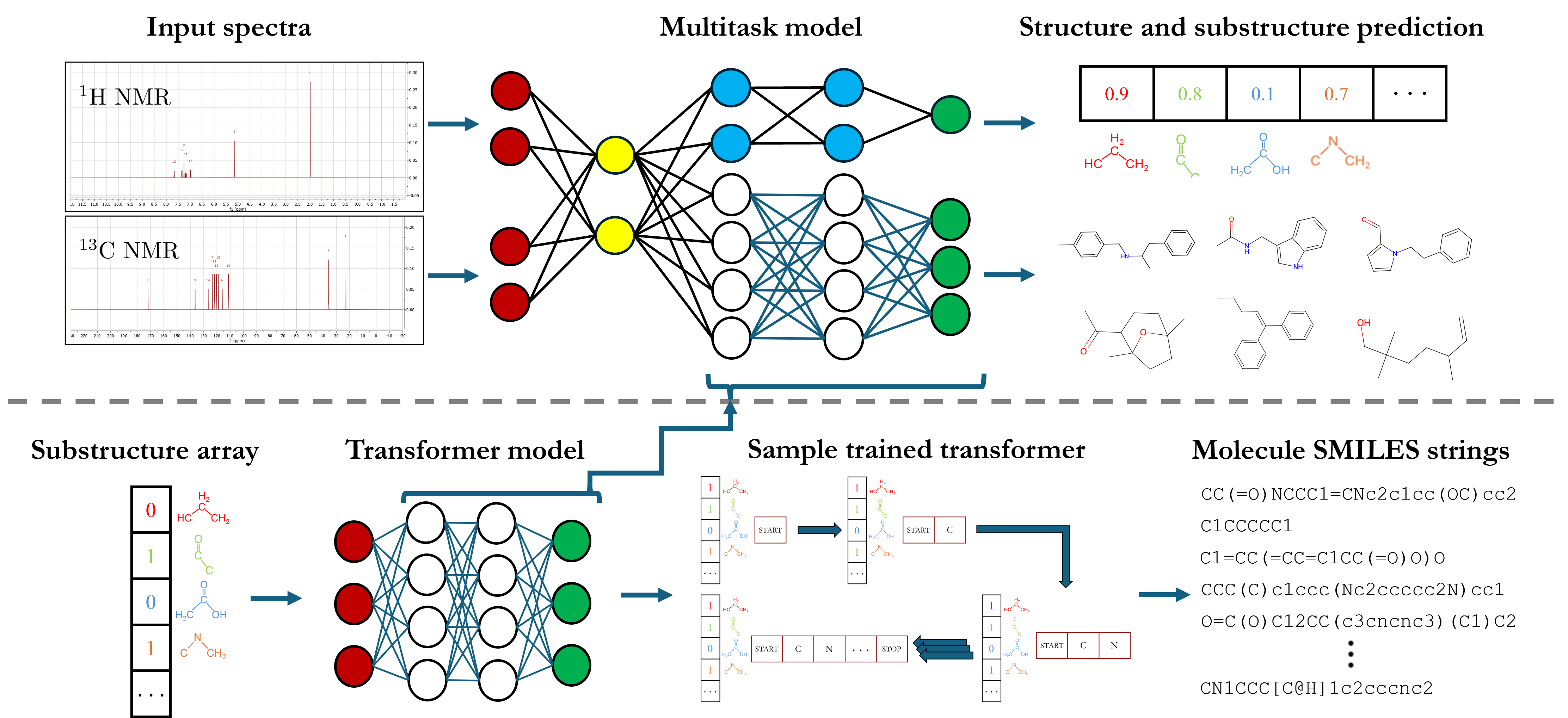}
    \caption{Overview of the full multitask structure elucidation workflow (top) and the substructure-to-structure workflow (bottom). Weights from a transformer pretrained on the substructure-to-structure task are used to initialize the multitask model. Specific details regarding the transformer model architecture and multitask model architecture can be found in SI Section~1.}
    \label{fig:overview}
\end{figure}

An overview of our structure elucidation framework is shown in Figure~\ref{fig:overview}. The top row of Figure \ref{fig:overview} illustrates our full end-to-end multitask structure elucidation model, which takes in the \textsuperscript{1}H and/or \textsuperscript{13}C NMR spectra and predicts both the substructures in the molecule and the molecule's structure. The bottom row of Figure~\ref{fig:overview} shows our approach to the substructure-to-structure problem, where molecular structures, represented by SMILES strings, are formed out of molecular fragments, represented by the substructure arrays. As we will show below, pre-training on the task of structure elucidation from substructures substantially improves the accuracy of the full multitask structure elucidation model. Hence, we first demonstrate how a transformer can be trained on this task before showing how this can be integrated into the complete end-to-end multitask framework.

\subsection{Substructure-to-structure prediction}\label{subsec:substruct_to_struct}
We first consider the problem of deducing a molecular structure from knowledge of the presence or absence of a set of substructures, which we define as small fragments of a molecule with defined bonding relationships. This task is inspired by the way chemists commonly interpret NMR spectra, using peaks to identify fragments and assembling those fragments into candidate structures.   Manually converting a relatively small number of substructures into a molecular structure is already challenging even for an experienced chemist, and it quickly becomes impractical when confronted with dozens or hundreds of substructures, which is necessary for accommodating the complexity of structure space. Hence, devising an automated strategy for structure elucidation from these molecular fragments can itself help accelerate structure determination.

Here, we treat the substructure-to-structure problem as a language translation problem, where we translate a sequence of substructures into a sequence of tokens that can be combined to form a molecule's SMILES string, recovering the molecular connectivity. Our approach is outlined in the bottom row of Figure \ref{fig:overview}, where we use a transformer model architecture for this task. Autoregressive transformer models have been shown to excel at translation tasks owing to the strong inductive bias provided by the multihead attention mechanism, which captures global correlations and uses that information to generate the target sequence in a conditioned manner\cite{vaswani_attention_2017}. This strength is critical in a chemical context because the overall connectivity of a molecule arises from considering all molecular fragments collectively and many fragments may or may not overlap depending on the molecular context. Here, we adapt a full encoder-decoder transformer architecture for the substructure-to-structure task.

The inputs to the substructure-to-structure transformer are a binary vector with each entry corresponding to the presence or absence of a specific substructure and a start token. Using only this information, the transformer generates a SMILES string token by token, which it continues to build until it terminates upon the prediction of the stop token.

To train the transformer on this task, we started with a set of $\sim$\numericalresult{143} thousand molecules from the SpectraBase\cite{john_wiley__sons_inc_spectrabase_nodate} database containing only C, N, and O as the non-hydrogen atoms and combined it with a set of $\sim$3 million molecules randomly sampled from the GDB-17 dataset\cite{ruddigkeit_enumeration_2012} to create a final dataset of $\sim$3.1 million molecules. After canonicalizing all the SMILES strings using RDKit\cite{noauthor_rdkit_nodate}, the substructure vector for each molecule was computed using RDKit's substructure match functionality against the set of 957 substructures used in our previous work\cite{huang_framework_2021} represented as SMART strings. For the target that the transformer uses in decoding, the canonicalized SMILES were tokenized using a regular expression\cite{schwaller_found_2018} and an alphabet was determined from all unique tokens across all SMILES to ensure that all SMILES can be represented during training and testing. Further details of our transformer training dataset are provided in SI Section~2.1.

\begin{figure}[h]
    \centering
    \includegraphics[width=\textwidth]{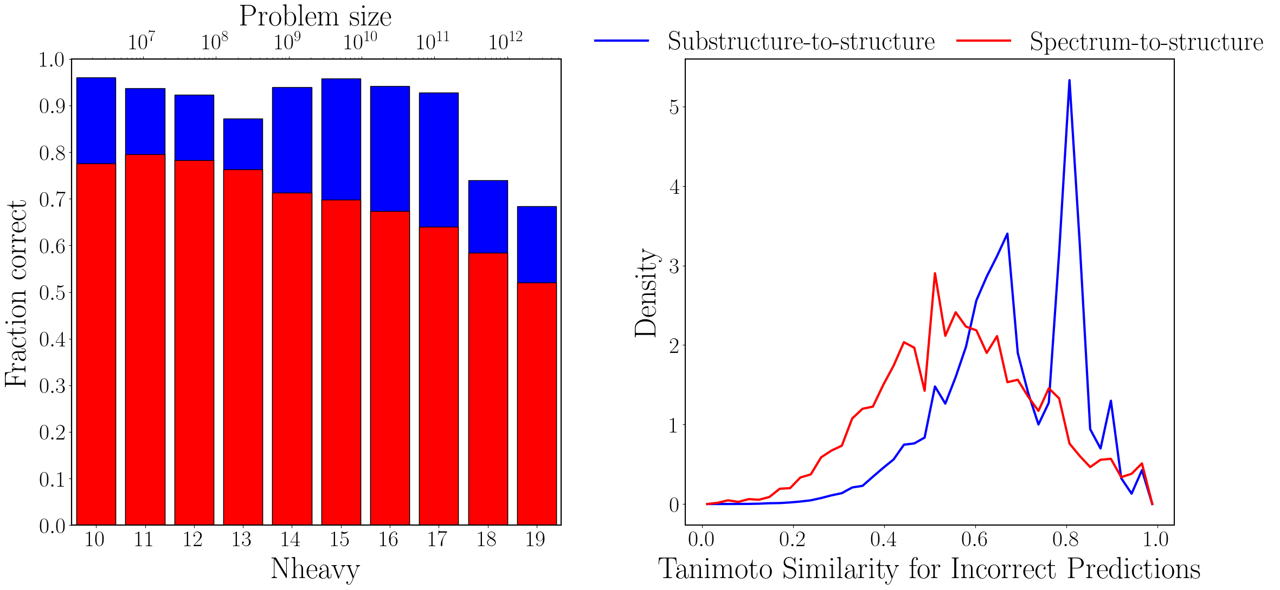}
    \caption{(Left) Transformer and the best multitask model test accuracy as a function of the problem size. The problem size is determined by extrapolating an exponential fit to the number of molecules in GDB-9\cite{ramakrishnan_quantum_2014}, GDB-11\cite{fink_virtual_2007}, GDB-13\cite{blum_970_2009}, and GDB-17, and the plot begins with the number of possible structures for 10 heavy (non-hydrogen) atoms. (Right) Distribution of Tanimoto similarities of the best incorrect predictions relative to the target molecule after removing correct predictions and invalid SMILES.}
    \label{fig:all_acc_per_size_and_tanimoto}
\end{figure}

During training, the dataset was split into a training, validation, and test set using a random split, with 80\% of the data used for training, 10\% used for validation, and 10\% used for testing. To produce a more compact representation of the binary substructure arrays for the embedding, an integer array was created which lists the index (1 to 957) of only the present (non-zero) substructures for each molecule. Each unique SMILES token was also assigned to an integer index and used for embedding. Right padding was applied to ensure all substructure vectors and SMILES vectors in a batch were of the same length before being passed into the transformer. The padding was masked during training. Further details of our transformer training procedure are provided in SI Section~3.1.

\begin{figure}[!ht]
    \centering
    \includegraphics[width=\textwidth]{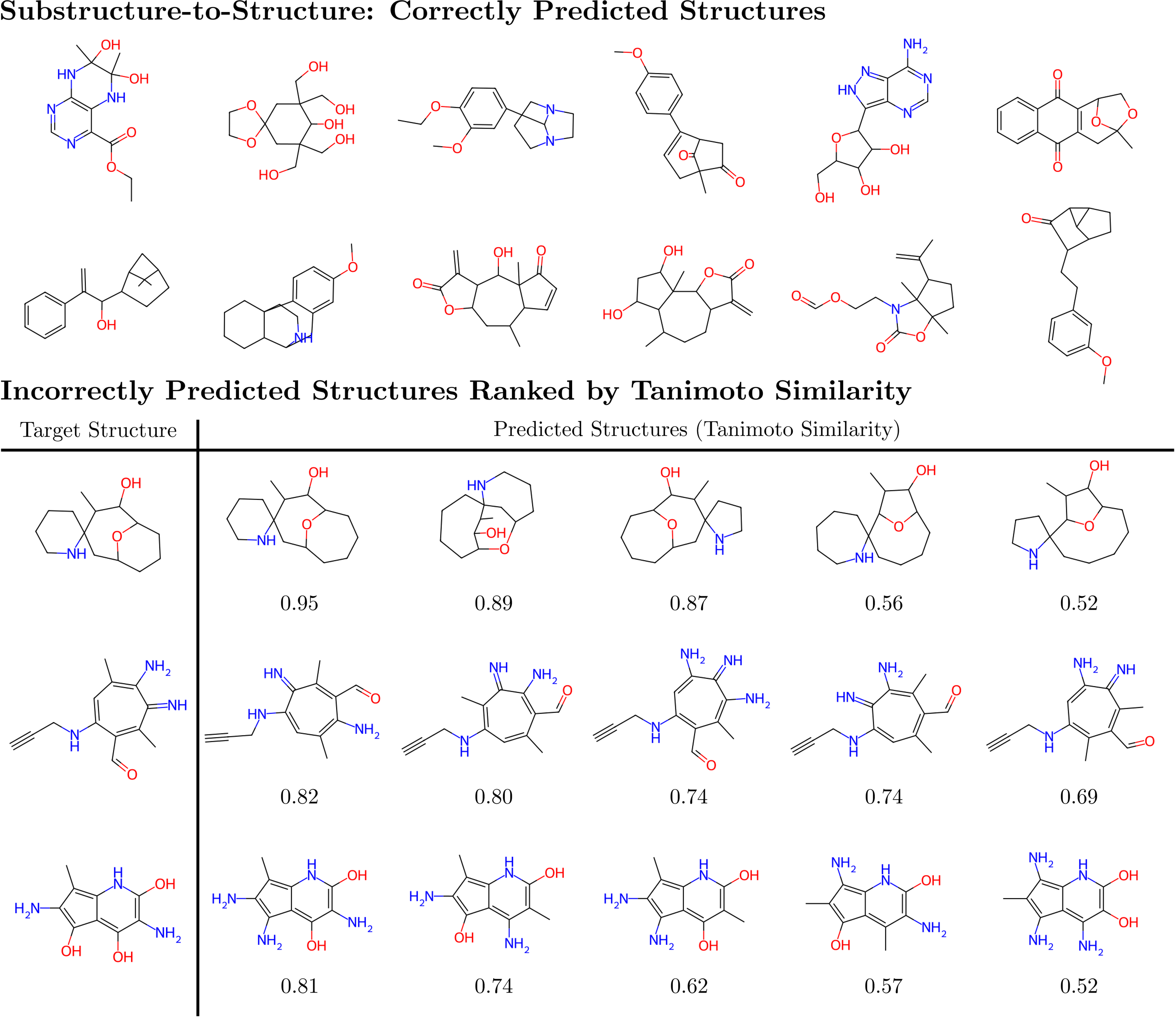}
    \caption{(Top) Examples of molecules that were correctly predicted by the transformer model. The molecules shown have between 23 to 65 substructures, and an example of a correctly predicted molecule and its constituent substructures is shown in SI Figure~4. (Bottom) Examples of molecules with incorrect predictions from the transformer model. The number beneath each predicted molecule is the Tanimoto similarity between the prediction and target.}
    \label{fig:transformer_tanimoto_examples}
\end{figure}

To test the trained transformer model for the substructure-to-structure task, top-$k$ random sampling\cite{fan_hierarchical_2018} was used to generate SMILES from a given substructure array with $k=5$, and 15 predictions were generated for each input. Using these $15$ generated SMILES, the prediction was considered correct if one of these molecules, upon canonicalization, matched the exact canonical SMILES of the actual molecule. On the test set, the model achieved an accuracy of \numericalresult{93.2\%}. Examples of molecules correctly predicted by the transformer model are shown at the top section of Figure~\ref{fig:transformer_tanimoto_examples}, demonstrating that the model can deduce complex molecular structures using only substructures as input. These example molecules require the transformer to assemble between 23 and 65 substructures into the correct molecule. The difficulty of this task is illustrated in SI Figure~4, which shows a molecule that was correctly predicted by the substructure-to-structure model and the 53 substructures that were provided to it. The test set accuracy as a function of the number of heavy atoms is shown as the blue data in the left panel of Figure \ref{fig:all_acc_per_size_and_tanimoto} for up to 19 heavy atoms. Notably, the accuracy of the substructure-to-structure model shows little variation as the problem size grows. For example, at 10 heavy atoms, where the problem size is $\sim$2 million, the accuracy is \numericalresult{96.0\%} and only drops to \numericalresult{92.8\%} by 17 heavy atoms, where it has increased to $\sim$200 billion. However, the accuracy drops below \numericalresult{80\%} for the molecules with 18 or 19 heavy atoms. Although this sudden decrease could be explained in part by the problem size entering the trillions for molecules of that size, a more likely explanation might be that GDB-17 only contains molecules with up to 17 heavy atoms, while the SpectraBase dataset, which does contain molecules with 18 and 19 heavy atoms, is much smaller. Hence, whereas \numericalresult{59\%} in our training data come from molecules with 17 heavy atoms, only \numericalresult{1\%} come from molecules with 18 or 19 heavy atoms, meaning that the model has had fewer opportunities to learn how to predict systems of that size. However, the fact that the model retains reasonable accuracy for these larger molecules suggests that the approach is generalizable. Since computing the substructure vectors is facile using RDKit, the set of substructures could also easily be expanded to cover fragments more commonly encountered in larger molecules, which would be expected to further improve performance.  

To better understand the failure modes of the substructure-to-structure model, we examine the distribution of Tanimoto similarities\cite{rogers_computer_1960} computed between the model's best incorrect prediction and the correct molecule in cases where the correct molecule was not predicted. This is shown for the test set as the blue line in the right panel of Figure ~\ref{fig:all_acc_per_size_and_tanimoto}. Overall, \numericalresult{89.6\%} of the best incorrect predictions have a similarity to the correct molecule that is greater than or equal to 0.50, with an average similarity of \numericalresult{0.68}. To see how these similarity scores correlate to the molecular structure of the incorrect predictions, the bottom section of Figure~\ref{fig:transformer_tanimoto_examples} shows some example target molecules and incorrect predictions at varying degrees of Tanimoto similarity. We see that even for incorrect predictions at similarities near 0.50, the predictions already contain many of the desired functional groups and bonding motifs of the target, albeit with incorrect connectivity. As the similarity increases, the connectivity and structure of the predictions improve, and at similarities near 0.80 and above, the incorrect predictions are very close to the target, often differing by a few atoms or bonds. This is encouraging as it shows that the transformer is capable of learning how to map the binary substructure representation to real molecular fragments, leading to the recovery of a significant portion of the target molecule scaffold even in cases where the exact correct molecule is not predicted. Considering both the high proportion of best incorrect predictions above 0.50 and the high mean similarity, these results show that even when the model returns incorrect predictions, the molecules generated can still provide meaningful structural information about the target molecule.

\subsection{Spectrum-to-structure and spectrum-to-substructure prediction using a multitask workflow}\label{subsec:spectrum_to_struct}

We now focus on the more challenging problem of directly elucidating the molecular structure and substructures from the \textsuperscript{1}H and \textsuperscript{13}C NMR spectra, which we refer to as spectrum-to-structure and spectrum-to-substructure prediction respectively. Our goal is to design a model capable of predicting a molecule's structure and the substructures it contains using only the spectral data as inputs, operating in an end-to-end fashion i.e. taking the spectrum and directly predicting the structure it corresponds to and the substructures within that molecule with no intermediate steps. The workflow for our approach is shown in the top row of Figure \ref{fig:overview}. As we will emphasize further below an important component of training this multitask spectrum-to-structure model is the initialization of the transformer using the weights obtained from training on the substructure-to-structure task (as indicated by the arrow in Figure \ref{fig:overview} from the bottom workflow to the multitask model on the top).

For training and testing the spectrum-to-structure model, we used \textsuperscript{1}H and/or \textsuperscript{13}C NMR spectra for the $\sim$\numericalresult{143} thousand set of molecules from SpectraBase predicted using MestreNova\cite{willcott_mestre_2009}, using the same train-validation-test split of the molecules as on the substructure-to-structure task. To effectively utilize all portions of the spectral signal, we took a minimal preprocessing approach to the spectra to retain as much information as possible. Specifically, since the \textsuperscript{1}H NMR contains detailed splittings and rapidly varying features, we processed the spectrum by mildly reducing the dimensionality of the data by linearly interpolating the spectrum of \numericalresult{32768} values down to a grid of 28000 values corresponding to a shift range of -2 to 12 ppm with a resolution of 0.0005 ppm. To assist the stability of the multitask model while retaining the relative intensities of all peaks, all intensities were normalized to between 0 and 1 by dividing each spectrum by the intensity of its highest peak. For the \textsuperscript{13}C NMR spectra, because the intensities are not generally reliable indicators of the relative number of nuclei and the peak shapes are not informative because peaks are typically decoupled, we processed the spectra by extracting the chemical shifts and binning them into 80 bins spanning the shift range of 3.42 to 231.3 ppm. Further details of our multitask model training dataset are provided in SI Section~2.2.

During training, we passed the processed \textsuperscript{1}H NMR spectra through a one-dimensional convolutional neural network (CNN) to downsample the signal into a lower-dimensional space. For the processed \textsuperscript{13}C NMR spectra, since it is a binary vector, we embedded it into a dense vector representation. If both spectra were being used as input to the spectrum-to-structure model then their features were concatenated. If only one spectrum was being used as input then only the features for that spectrum were retained. These features were then passed through a full encoder-decoder transformer for structure elucidation, which outputs SMILES strings, and a transformer encoder for substructure elucidation, which outputs substructure probability arrays as shown schematically on the top row of Figure~\ref{fig:overview}. Further details of our multitask model training procedure are provided in SI Section~3.2.

Table~\ref{table:multitask_model_performance} shows the structure elucidation accuracy on the test set as a function of both the spectral data used (\textsuperscript{1}H and/or \textsuperscript{13}C NMR) and if weights from a transformer pretrained on the substructure-to-structure task were used, in other cases randomly initialized weights were used. Structure elucidation accuracy was tested in the same way as for the substructure-to-structure task, where a target was considered correctly predicted if its canonical SMILES string appeared within the set of 15 predictions upon canonicalization. The highest accuracy of \numericalresult{69.6\%} (Table~\ref{table:multitask_model_performance}, third row) is obtained when using both the \textsuperscript{1}H and \textsuperscript{13}C NMR spectra with a pretrained transformer for structure elucidation. Without using the pretrained transformer, the structure elucidation accuracy drops significantly from \numericalresult{69.6\%} to \numericalresult{53.3\%}, demonstrating the importance of the pretraining step to the success of the overall multitask framework. This result suggests that pre-training on the substructure-to-structure task helps the model learn a chemically relevant latent space that translates well when integrated into the multitask framework.

To investigate the relative importance of the \textsuperscript{1}H and \textsuperscript{13}C NMR as inputs, we trained two versions of the model using either \textsuperscript{1}H or \textsuperscript{13}C NMR as the sole input. The accuracy of the model is significantly better with only \textsuperscript{1}H NMR (\numericalresult{59.6\%}) than with only \textsuperscript{13}C NMR (\numericalresult{22.0\%}), reflecting the fact that \textsuperscript{1}H NMR contains far more information about the molecule's connectivity relative to \textsuperscript{13}C NMR, thus playing a more critical role in structure elucidation. However, the best accuracy is only obtained when combining both spectra, showing that they are complementary to each other in terms of the information they each provide. We also note that using \textsuperscript{1}H NMR alone with the pretrained transformer gives a greater structure prediction accuracy than obtained using both \textsuperscript{1}H and \textsuperscript{13}C NMR data without the pretrained transformer, highlighting further the advantages of this approach.

\begin{figure}[h]
    \centering
    \includegraphics[width=\textwidth]{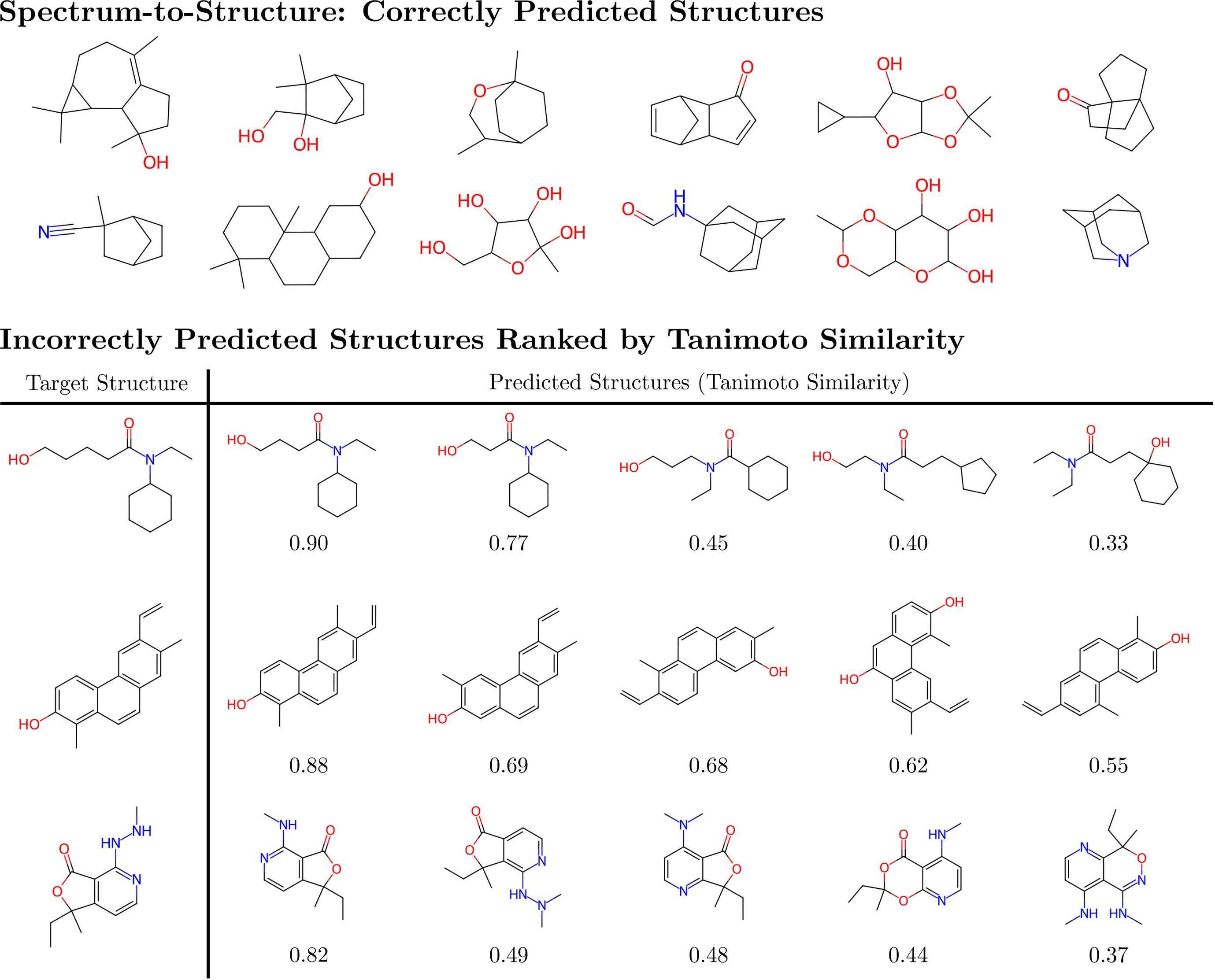}
    \caption{(Top) Examples of molecules that were correctly predicted using the multitask spectrum-to-structure model. (Bottom) Examples of molecules with incorrect predictions from the multitask model. The number beneath each predicted molecule is the Tanimoto similarity between the prediction and target.}
    \label{fig:multitask_tanimoto_examples}
\end{figure}

The left panel of Figure \ref{fig:all_acc_per_size_and_tanimoto} shows in red the structure elucidation accuracy of the spectrum-to-structure multitask model as a function of the number of heavy atoms. In comparison to the substructure-to-structure model there is a more systematic decrease in accuracy with the size of the molecule, decreasing from \numericalresult{77.5\%} at 10 heavy atoms to \numericalresult{52.0\%} at 19 heavy atoms. However, this decrease in accuracy is dwarfed by the increase in the problem size that grows combinatorially and thus increases over 5 orders of magnitude over that range of molecule sizes. Hence, not only does the model's accuracy decay extremely slowly compared to the growth of the number of possible structures, it is able to achieve this impressive scaling without any reliance on the molecular formula, molecular weight, or other information about the system that would otherwise constrain the problem. The top section of Figure~\ref{fig:multitask_tanimoto_examples} shows some of the molecules that were correctly predicted, emphasizing that the multitask model can elucidate the structures of molecules solely from their \textsuperscript{1}H and \textsuperscript{13}C NMR spectra across a wide range of chemical motifs that would be extremely challenging without additional information.

\begin{table}[!ht]
    \centering
    \caption{Test set structure elucidation accuracy and substructure elucidation $F_1$ score of the multitask model as a function of the type of data used and weight initialization for the transformer component.}
    \label{table:multitask_model_performance}
    \sisetup{round-mode=places}
    \large
    \resizebox{\columnwidth}{!}{\begin{tabular}{ccS[round-precision=1]c}
        \toprule
        \textbf{Data Used} & \textbf{Pretrained Transformer} & \textbf{Structure Accuracy (\%)} & \textbf{Substructure $F_1$ Score} \\ \midrule
        \textsuperscript{13}C NMR Only & Yes & \numericalresult{22.0} & \numericalresult{0.69}\\ 
        \textsuperscript{1}H NMR Only & Yes & \numericalresult{59.6} & \numericalresult{0.81} \\ 
        \textsuperscript{1}H + \textsuperscript{13}C NMR & Yes & \numericalresult{69.6} & \numericalresult{0.86} \\ 
        \textsuperscript{1}H + \textsuperscript{13}C NMR & No & \numericalresult{53.3} & \numericalresult{0.86} \\ \midrule
    \end{tabular}}
\end{table}

For the molecules where the spectrum-to-structure model was unable to predict the correct molecule, we can again use Tanimoto similarities to assess how close the predicted molecule was. The distribution of Tanimoto similarities of the best incorrect predictions for the spectrum-to-structure task is shown in red in the right panel of Figure \ref{fig:all_acc_per_size_and_tanimoto}. From this we see that compared to the substructure-to-structure task, the distribution of similarity scores is shifted to lower values, with only \numericalresult{64.7\%} of incorrect predictions having a similarity of 0.50 or greater with a mean similarity of \numericalresult{0.56}. The lower similarity scores relative to the substructure-to-structure task reflect the fact that direct structure elucidation from only the NMR spectra is a considerably more challenging inverse problem that is less constrained than constructing the structure from a set of substructures. However, the bottom section of Figure~\ref{fig:multitask_tanimoto_examples} shows that the molecules with the highest Tanimoto similarity obtained from incorrect predictions of the multitask model still contain many of the chemical motifs expected in the target compound, and so still provide insight into the system that is valuable when deducing the molecular connectivity.

The multitask model also outputs a prediction for the target molecule's substructure profile, which can be interpreted as the probability of a given molecular fragment being present in the molecule based on its NMR spectra. These profiles, which provide additional information as to which of the 957 substructures are present in the system, are useful in cases where the model does not arrive at the correct structure since they could be used by a chemist to infer other possible structures of the molecule. To evaluate the substructure elucidation accuracy, one must be careful to take into account that most molecules only contain a relatively small fraction of the total 957 substructures and therefore the profiles are dominated by 0's (which indicate the absence of a substructure). This leads to a highly imbalanced classification problem where it is insufficient to only use accuracy as a performance metric. To address this issue Table~\ref{table:multitask_model_performance} shows the $F_1$ score that balances the need to account for true and false positives as well as negatives. From this, we see that for this spectrum-to-substructure task, while the performance improvements in the $F_1$ score of using both \textsuperscript{1}H and \textsuperscript{13}C NMR spectra over either individually are retained, the pretraining of the transformer has a negligible impact on the substructure elucidation. This is perhaps expected since although the multitask model produces both a prediction of structure and substructures in an end-to-end fashion, the pretrained transformer weights arise from the task of substructure-to-structure prediction, which is a downstream task from substructure prediction.

\begin{figure}[!ht]
    \centering
    \includegraphics[width=\textwidth]{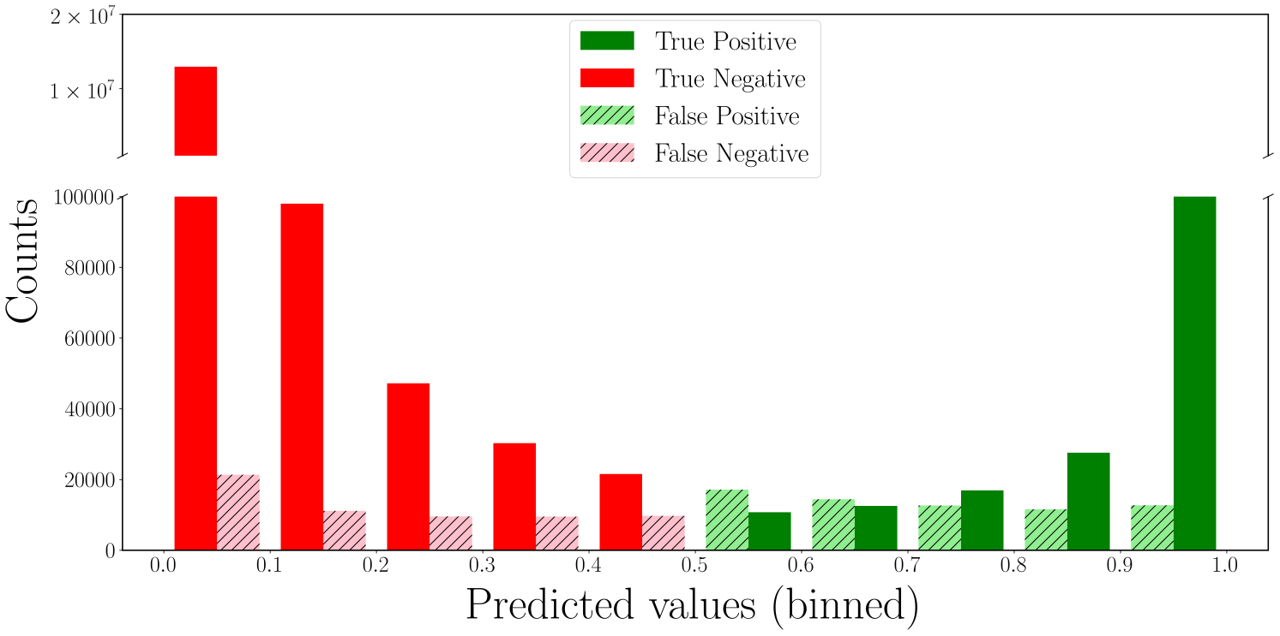}
    \caption{Distribution of true/false positives and true/false negatives as a function of the probability predicted by the multitask model using both \textsuperscript{1}H and \textsuperscript{13}C NMR as inputs and a pretrained transformer. A decision boundary of 0.5 is used to distinguish between positives and negatives.}
    \label{fig:multitask_substruct_dist}
\end{figure}

To quantify the model's predictive performance on the spectrum-to-substructure task beyond measures like the $F_1$ score, we can examine the distribution of the model's predicted probabilities. Figure \ref{fig:multitask_substruct_dist} shows the distribution of model predictions on the test set. In cases where the model predicts a low probability of a substructure being present ($<$0.1), the prediction is \numericalresult{99.8\%} accurate. Conversely, when the model predicts a high probability of a substructure being present ($>$0.9), the prediction is \numericalresult{96.3\%} accurate. The slightly lower accuracy in the case of predicting positive substructures arises from the significant imbalance of positives to negatives since in the training data each molecule only contains a small fraction of the 957 substructures, and so there are far more true negatives than true positives in the dataset. The uncertainty of the model increases when moving away from the extremes of the predicted probability towards the decision boundary of 0.5 between positive and negative predictions; however, only \numericalresult{2.6\%} of substructure predictions have a probability within the range of $0.1<p<0.9$, so the predicted probability from the model is a strong indicator of the presence or absence of a substructure the majority of the time.

\section{Conclusion}

In summary, we have introduced a multitask machine learning approach for direct structure and substructure elucidation from \textsuperscript{1}H and \textsuperscript{13}C NMR spectra that leverages pretraining an encoder-decoder transformer on the related task of substructure-to-structure elucidation. By integrating this transformer architecture into our multitask framework, we have shown that our end-to-end model is able to predict the structure correctly \numericalresult{69.6\%} within the first 15 predictions using only the \textsuperscript{1}H and \textsuperscript{13}C NMR as input for systems of up to 19 heavy atoms. Furthermore, our end-to-end model can simultaneously predict which substructures are present in a molecule with an accuracy of \numericalresult{96.3\%} when the predicted probability is above 0.9 and which substructures are not present with \numericalresult{99.8\%} accuracy when the predicted probability is below 0.1, with only \numericalresult{2.6\%} of the predicted substructure probabilities not falling within those ranges. What is particularly remarkable is that while the problem size (number of possible molecules that can be constructed consistent with basic chemical bonding rules) grows combinatorially with the number of heavy atoms from 10 to 19 heavy atoms our multitask model shows only a \numericalresult{25.5\%} decrease in accuracy over a range in which the number of possible molecules increases by \numericalresult{5} orders of magnitude. This suggests that our approach to the inverse problem of spectrum-to-structure is scalable to larger chemical systems. Our model thus provides an efficient avenue for direct spectrum-to-structure elucidation from \textsuperscript{1}H and \textsuperscript{13}C NMR spectra without dependence on any prior chemical information such as the molecular formula or molecular fragments.

This work sets the stage for future developments of this multitask framework to elucidate even larger molecules with an extended range of elements and the prediction of stereochemistry. Enabling these developments will require expanding the set of substructures used to pretrain the transformer with substructures containing additional elements and larger, more complicated chemical motifs such as larger ring systems or protecting groups. Relative stereochemistry could be incorporated within our existing multitask framework by training on SMILES representations containing the characters specifying stereocenters and \textit{cis-} and \textit{trans-}double bond configurations and the corresponding NMR spectra of molecules. This could be achieved by identifying all the stereocenters in the current training set and enumerating all possible stereoisomers and using these to augment the training set. An alternative approach to tackling the stereochemistry problem is to predict composition and connectivity as shown here and then do forward prediction of the spectra of different stereoisomers of candidate structures and compare to the input spectra to identify the most likely stereoisomer. 

Although in this work we have concentrated on absolute structure prediction with no chemical knowledge of the compound beyond its NMR spectra, this framework could also be easily adapted for cases where some information about the chemical system is available, such as in the case of reaction prediction or retrosynthesis where the starting materials of a reaction are known, and hence the problem is considerably more constrained. We also envision this technology to be used in a complementary manner with other existing structure elucidation, reaction prediction, or retrosynthesis frameworks as a way to rapidly detect structural or substructural changes in a reaction pathway or as a way to quickly constrain the number of possible candidate structures to a point where more refined techniques can be applied tractable. To enable such adaptations and future developments, we have made the code available on GitHub\cite{frank_hu_nmr2struct_2024}. 
Our approach opens up new possibilities in the field of ML-driven structure elucidation by introducing a fast and efficient structure elucidation framework that can operate in an unsupervised manner without relying on additional knowledge. The trained model, which is provided in our GitHub repository\cite{frank_hu_nmr2struct_2024}, can make a full prediction of the structure and substructures of an input \textsuperscript{1}H and \textsuperscript{13}C NMR spectra for a system of 19 heavy atoms in under 3 seconds even on standard CPU hardware (a single core of an AMD Ryzen 7 3700X 8-core processor). This framework thus has the potential to provide a highly accessible technology to greatly accelerate characterization and chemical discovery at levels ranging from high school chemical education to industrial research settings. Ultimately, we envision a community-driven effort to aid in curating a large database of experimental NMR spectra, which will help the model learn to predict structure from spectra collected under a variety of conditions. This synergistic effort of assisting community chemical characterization efforts in tandem with improving the prediction model provides an opportunity to greatly improve the accuracy of this tool while supporting the chemistry community.

\begin{suppinfo}
The supporting information contains the architectures and hyperparameters for all models presented in the main text, details on the data curation for the substructure-to-structure, spectrum-to-structure, and spectrum-to-substructure tasks, optimization protocol and learning curves for all models presented in the main text, and additional examples of correctly predicted molecules with the associated spectra or substructures.
\end{suppinfo}

\begin{acknowledgement}
F.H. acknowledges support from a Stanford Center for Molecular Analysis and Design (CMAD) fellowship. Some of the computing for this project was performed on the Sherlock cluster. We would like to thank Stanford University and the Stanford Research Computing Center for providing computational resources and support that contributed to these research results. The authors acknowledge Shriram Chennakesavalu for helpful discussions about transformers.
\end{acknowledgement}

\bibliography{references}

\clearpage
\noindent For Table of Contents only:

\includegraphics[width=1.0\columnwidth]{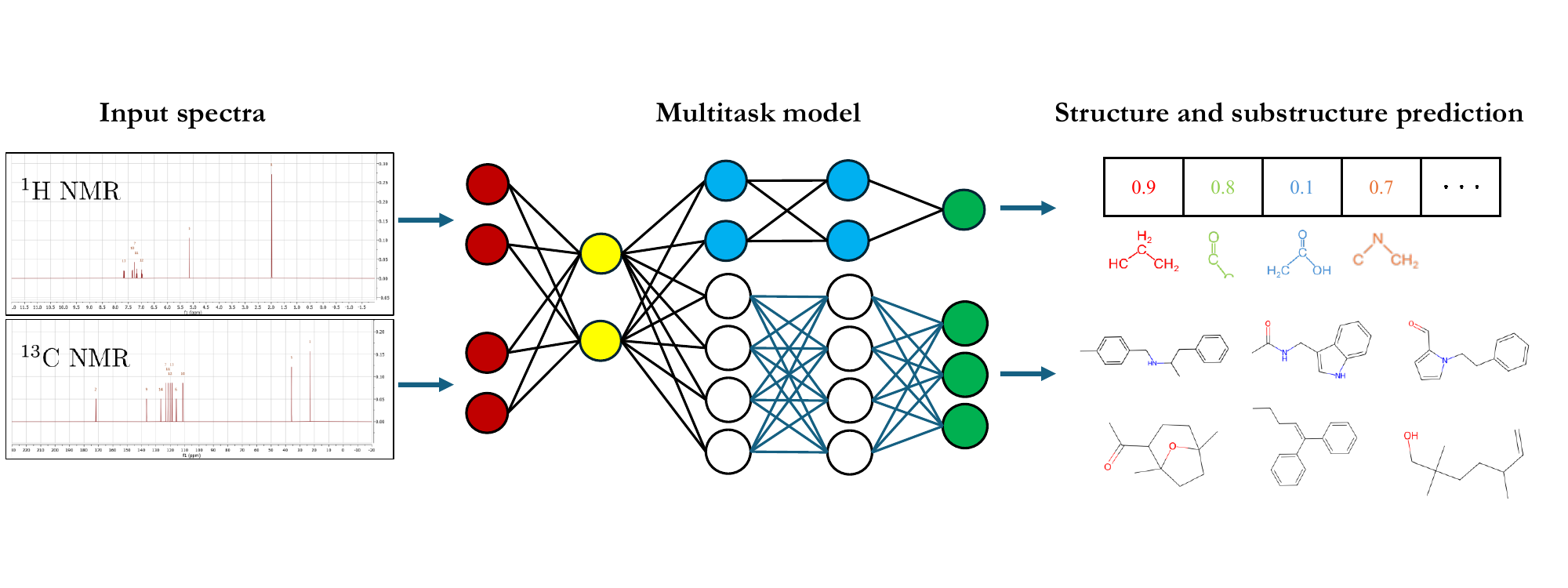}

\noindent Synopsis: 

We introduce a multitask machine learning framework that rapidly predicts both the molecular structure and molecular fragments of an unknown compound using only one-dimensional \textsuperscript{1}H and \textsuperscript{13}C NMR spectra.

\end{document}


\maketitle
\tableofcontents
\newpage
\section{Model architectures}\label{sec:SI_mod_architecture}
\begin{figure}
    \centering
    \includegraphics[width=\textwidth]{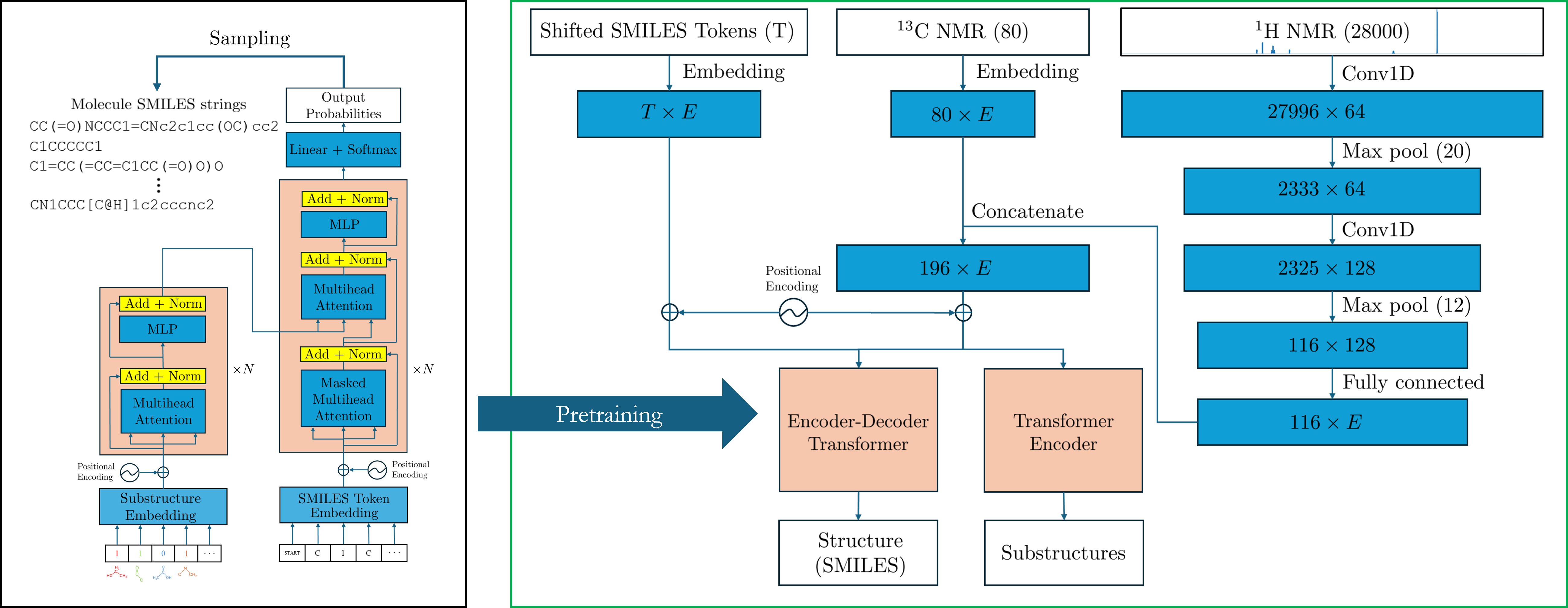}
    \caption{Diagram of the multitask workflow during training, with the substructure-to-structure transformer (left, black box) feeding into the overall multitask architecture (right, green box). $E$ refers to the model dimension, $T$ is the length of the target SMILES string in tokens, and $N$ is the number of encoder/decoder layers in the transformer. The transformer (left, black box) uses the PyTorch Transformer class. Conv1D is a one-dimensional convolution using the PyTorch Conv1d layer, MLP (a multi-layer perceptron) refers to a feed-forward neural network as described in the original transformer paper\cite{vaswani_attention_2017}, max pooling is performed using the PyTorch MaxPool1d layer, embedding is done using the PyTorch Embedding layer, and Norm refers to layer normalization.}
    \label{fig:SI_mod_architecture}
\end{figure}

For the substructure-to-structure transformer (Figure \ref{fig:SI_mod_architecture}, black box), we use an encoder-decoder transformer\cite{vaswani_attention_2017} as implemented in PyTorch\cite{paszke_pytorch_2019}. Table \ref{table:SI_substruct_to_struct_transformer_architecture} details the exact architecture of the final model used for the results in the main text. All models used absolute positional encoding based on sinusoidal functions of different frequencies\cite{vaswani_attention_2017} as described by the following equations:
\begin{align*}
    PE_{(pos, 2i)}   &= \sin(pos/10000^{2i/d_{model}})\\
    PE_{(pos, 2i+1)} &= \cos(pos/10000^{2i/d_{model}})
\end{align*}
where $pos$ is the position, $i$ is the dimension, and $d_{model}$ is the model embedding dimension. Positional encodings are added to the embeddings for both the substructures and the SMILES tokens.

\begin{table}[!ht]
    \centering
    \caption{Substructure-to-structure architectural parameters, descriptions, and their values.}
    \label{table:SI_substruct_to_struct_transformer_architecture}
    \resizebox{\columnwidth}{!}{\begin{tabular}{lll}
        \toprule
        \textbf{Parameter} & \textbf{Description} & \textbf{Value} \\ \midrule
        d\_model & Embedding dimension of the model & 128\\ 
        dim\_feedforward & Hidden layer dimension of the feed forward neural network within the transformer blocks & 1024 \\
        source\_size & Total number of possible token values for embedding substructure sequences & 958 \\
        src\_pad\_token & The index used for padding source sequences to the same length & 0\\
        target\_size & Total number of possible token values for embedding SMILES token sequences & 24\\
        tgt\_pad\_token & The index used for padding target sequences to the same length & 21 \\ 
        num\_encoder\_layers & The number of encoder layers & 6\\
        num\_decoder\_layers & The number of decoder layers & 6\\
        nhead & The number of heads used in multihead attention & 8\\
        activation & The activation function used for intermediate encoder/decoder layers & relu\\
        dropout & The probability for a particular element of an input tensor to be randomly set to 0 & 0.1\\
        layer\_norm\_eps & The constant used for numerical stability in layer normalization\cite{ba_layer_2016} & 1E-5\\ \midrule
    \end{tabular}}
\end{table}

\begin{table}[!ht]
    \centering
    \caption{Spectrum-to-substructure encoder architectural parameters, descriptions, and their values.}
    \label{table:SI_multitask_encoder_architecture}
    \resizebox{\columnwidth}{!}{\begin{tabular}{lll}
        \toprule
        \textbf{Parameter} & \textbf{Description} & \textbf{Value} \\ \midrule
        d\_model & Embedding dimension of the model & 128\\ 
        dim\_feedforward & Hidden layer dimension of the feed forward neural network within the transformer blocks & 1024 \\
        num\_encoder\_layers & The number of encoder layers & 4\\
        nhead & The number of heads used in multihead attention & 4\\
        activation & The activation function used for intermediate encoder/decoder layers & relu\\
        dropout & The probability for a particular element of an input tensor to be randomly set to 0 & 0.1\\
        layer\_norm\_eps & The constant used for numerical stability in layer normalization\cite{ba_layer_2016} & 1E-5\\ \midrule
    \end{tabular}}
\end{table}

For the spectrum-to-structure and spectrum-to-substructure multitask model, the architecture of the convolutional embedding for the \textsuperscript{1}H NMR and embedding for the \textsuperscript{13}C NMR is shown on the right side in the green box of Figure \ref{fig:SI_mod_architecture}. The architecture of the encoder-decoder transformer component is the same as in Table \ref{table:SI_substruct_to_struct_transformer_architecture}, and Table \ref{table:SI_multitask_encoder_architecture} describes the architecture of the encoder component used for substructure elucidation in the final multitask model. An important step in obtaining the substructure profile from the encoder is aggregating the information from the higher-dimensional raw encoder output into the lower-dimensional format of the substructure profiles. To this end, we adapt a method called sequence pooling\cite{hassani_escaping_2022}. Given an input sequence $\mathbf{x}_0$ and a function $f$ parameterized by a neural network, sequence pooling performs the following operations in order:
\begin{align}
    \mathbf{x}_L &= f(\mathbf{x}_0)\in\mathbb{R}^{N\times T \times E}\\
    \mathbf{x}'_L &= \text{softmax}(g(\mathbf{x}_L)^T)\in \mathbb{R}^{N \times 1 \times T}\\
    \mathbf{z} &= \text{squeeze}(\mathbf{x}'_L\mathbf{x}_L)\in \mathbb{R}^{N\times E}
\end{align}
where $N$ is the batch size, $T$ is the sequence length, $E$ is the embedding dimension, and $g(\cdot)\in \mathbb{R}^{E\times 1}$ is a learnable linear transformation. This can be understood as attending across the sequence dimension of the data after processing it through the encoder and assigning importance weights to each element in the sequence before aggregation. 

\section{Data curation}\label{sec:SI_data_curation}

\subsection{Substructure-to-structure transformer}\label{subsec:SI_data_substruct_to_struct}
To generate the data for training the substructure-to-structure transformer model, we started with a set of 142894 SMILES strings containing only C, N, O, and H atoms from the SpectraBase\cite{john_wiley__sons_inc_spectrabase_nodate} dataset. We canonicalized all the SMILES strings using RDKit\cite{noauthor_rdkit_nodate} and removed all stereochemical information from the strings, including designation of double bond stereochemistry. We combined this set of 142894 SMILES with randomly sampled SMILES from the GDB-17 dataset\cite{ruddigkeit_enumeration_2012} to create a final dataset of 3116791 SMILES strings. Using the set of 957 substructures from our previous work\cite{huang_framework_2021}, we constructed the substructure arrays for this dataset by performing a substructure search, generating a binary vector of length 957 for each molecule where a ``1'' means a substructure is present and a ``0'' means a substructure is absent. SMILES strings were tokenized using a regular expression\cite{schwaller_found_2018}.

\subsection{Spectrum-to-structure and spectrum-to-substructure multitask model}\label{subsec:SI_data_spectrum_to_struct}
To generate data for training the spectrum-to-structure and spectrum-to-substructure multitask model, we converted the set of 142894 SMILES strings from SpectraBase into two-dimensional mol files using open babel\cite{oboyle_open_2011} and then computed the \textsuperscript{1}H and \textsuperscript{13}C NMR spectra using MestreNova\cite{willcott_mestre_2009} version 14.2. 

\textsuperscript{1}H NMR spectra were computed with a line width of 0.75 Hz using deuterated chloroform as the solvent. Labile protons were excluded from the calculation of the spectra. The spectra were output as grids of 32768 intensities which were then interpolated down to grids of 28000 values spanning a ppm shift range of -2 to 12 ppm with a resolution of 0.0005 ppm. Each spectrum was normalized by dividing by the intensity of its highest peak, thereby normalizing all intensities to between 0 and 1 to assist with model stability. \textsuperscript{13}C NMR spectra were computed with a line width of 1.50 Hz in proton decoupled mode. Chemical shifts were extracted from the spectrum and binned into 80 bins spanning the shift range of 3.42 to 231.3 ppm.

\section{Optimization protocol}\label{sec:SI_opt_protocol}

\subsection{Substructure-to-structure transformer}\label{subsec:SI_transformer_opt}
\begin{figure}
    \centering
    \includegraphics[width=0.6\textwidth]{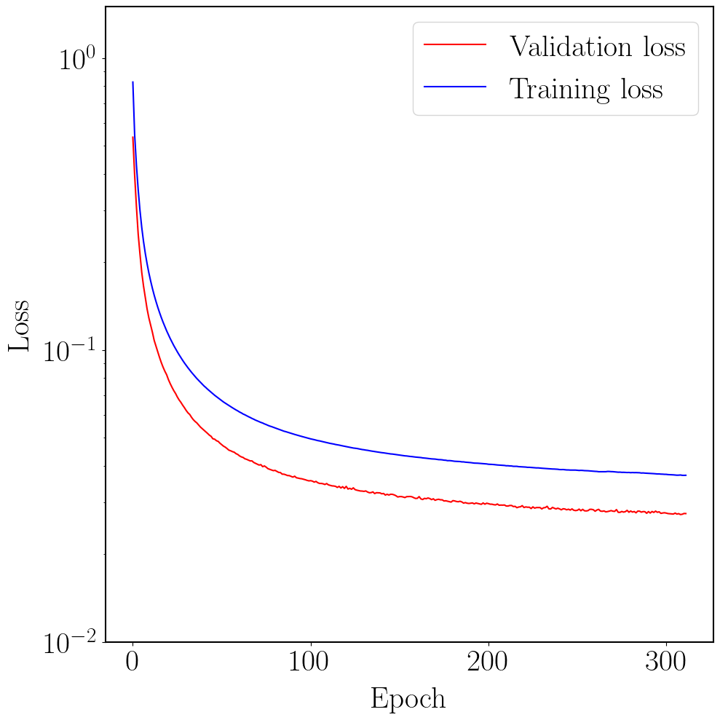}
    \caption{Training and validation loss curves for the transformer model used in the main text.}
    \label{fig:SI_transformer_learning_curve}
\end{figure}

The substructure-to-structure transformer model was trained using a batch size of 32 with a constant learning rate of $1\times 10^{-5}$ and a weight decay of $1\times 10^{-5}$ using the ADAM\cite{kingma_adam_2017} optimizer implemented in PyTorch. The loss function for the substructure-to-structure model was a cross entropy loss between the model's output probability distribution and the correct sequence of tokens for the target SMILES string. The dataset composed of the 3116791 SMILES strings and substructure arrays was partitioned using a random splitting with 80\% used for training, 10\% used for validation, and 10\% used for testing. Early stopping was used to prevent overfitting by monitoring the loss on the validation set. The model was optimized for \numericalresult{324} epochs and the final model selected was that with the lowest validation loss. This model was the one used to compute the results shown in the main text. The final substructure-to-structure model was trained for 308 epochs, at which point the minimum validation loss was attained. The learning curve for the training of the substructure-to-structure model is shown in Fig.~\ref{fig:SI_transformer_learning_curve}.

\subsection{Spectrum-to-structure and spectrum-to-substructure multitask model}\label{subsec:SI_multitask_opt}
The four spectrum-to-structure and spectrum-to-substructure multitask models, which correspond to different conditions of input data and whether a pretrained transformed was used, shown in Table~1 of the main text were trained using the following procedure. The models were trained using a constant learning rate of $1\times 10^{-5}$ and a batch size of 64 without any weight decay. The loss function being optimized for each multitask model is a weighted sum of the cross entropy loss for the SMILES prediction and a binary cross entropy loss for the substructure prediction. Given a prediction-target pair for the SMILES prediction $(\boldsymbol{y}_{smi}, \hat{\boldsymbol{y}}_{smi})$ and a prediction-target pair for the substructure prediction $(\boldsymbol{y}_{sub}, \hat{\boldsymbol{y}}_{sub})$, the total loss is computed as:
\begin{align*}
    \mathcal{L}_{tot} = \alpha \cdot\mathcal{L}_{CE}(\boldsymbol{y}_{smi}, \hat{\boldsymbol{y}}_{smi}) + \beta \cdot\mathcal{L}_{BCE}(\boldsymbol{y}_{sub}, \hat{\boldsymbol{y}}_{sub})
\end{align*}
where $\alpha$ and $\beta$ were set to $\alpha=\beta=1$.

\begin{figure}[ht]
    \centering
    \includegraphics[width=\textwidth]{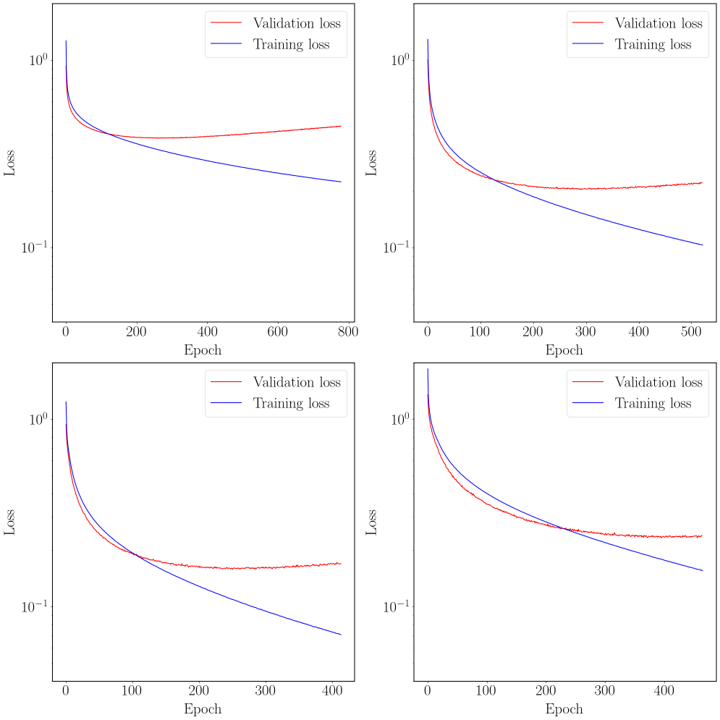}
    \caption{Training and validation loss curves for the multitask models used in the main text: (top left) \textsuperscript{13}C NMR only with pretrained transformer, (top right) \textsuperscript{1}H NMR only with pretrained transformer, (bottom left) \textsuperscript{13}C + \textsuperscript{1}H NMR with pretrained transformer, and (bottom right) \textsuperscript{13}C + \textsuperscript{1}H NMR without pretrained transformer.}
    \label{fig:SI_multitask_learning_curve}
\end{figure}

The dataset of 142894 SMILES strings, substructure arrays, and spectra was partitioned using the same train-validation-test split as in the substructure-to-structure task. For each of the four models early stopping was used to prevent overfitting by monitoring the loss on the validation set. The total number of epochs that each of the four models was trained for is shown in Table~\ref{table:SI_multitask_convergence}. The final model selected for each set of conditions was that with the lowest validation loss with the number of epochs at which that loss was reached shown in Table~\ref{table:SI_multitask_convergence} for each of the four models. The learning curves for the training of the spectrum-to-structure and spectrum-to-substructure multitask models are shown in Fig.~\ref{fig:SI_multitask_learning_curve}. By comparing the learning curves we note that using a pretrained transformer not only improves the structure elucidation accuracy, as described in the main text, but also accelerates convergence relative to training the multitask model from scratch. 

\begin{table}[!ht]
    \centering
    \caption{Number of epochs each multitask model was trained to attain the minimum validation loss.}
    \label{table:SI_multitask_convergence}
    \resizebox{\columnwidth}{!}{\begin{tabular}{llll}
        \toprule
        \textbf{Data used} & \textbf{Pretrained Transformer} &  \textbf{Total number of epochs trained} & \textbf{Final model epoch} \\ \midrule
        \textsuperscript{13}C NMR Only & Yes & 779 & 261\\
        \textsuperscript{1}H NMR Only & Yes & 521 & 288 \\
        \textsuperscript{1}H + \textsuperscript{13}C NMR & Yes & 414  & 250\\
        \textsuperscript{1}H + \textsuperscript{13}C NMR & No & 465 & 456\\ \midrule
    \end{tabular}}
\end{table}
\newpage
\newgeometry{textheight=240mm}
\begin{figure}[ht]
    \centering
    \includegraphics[width=\textwidth]{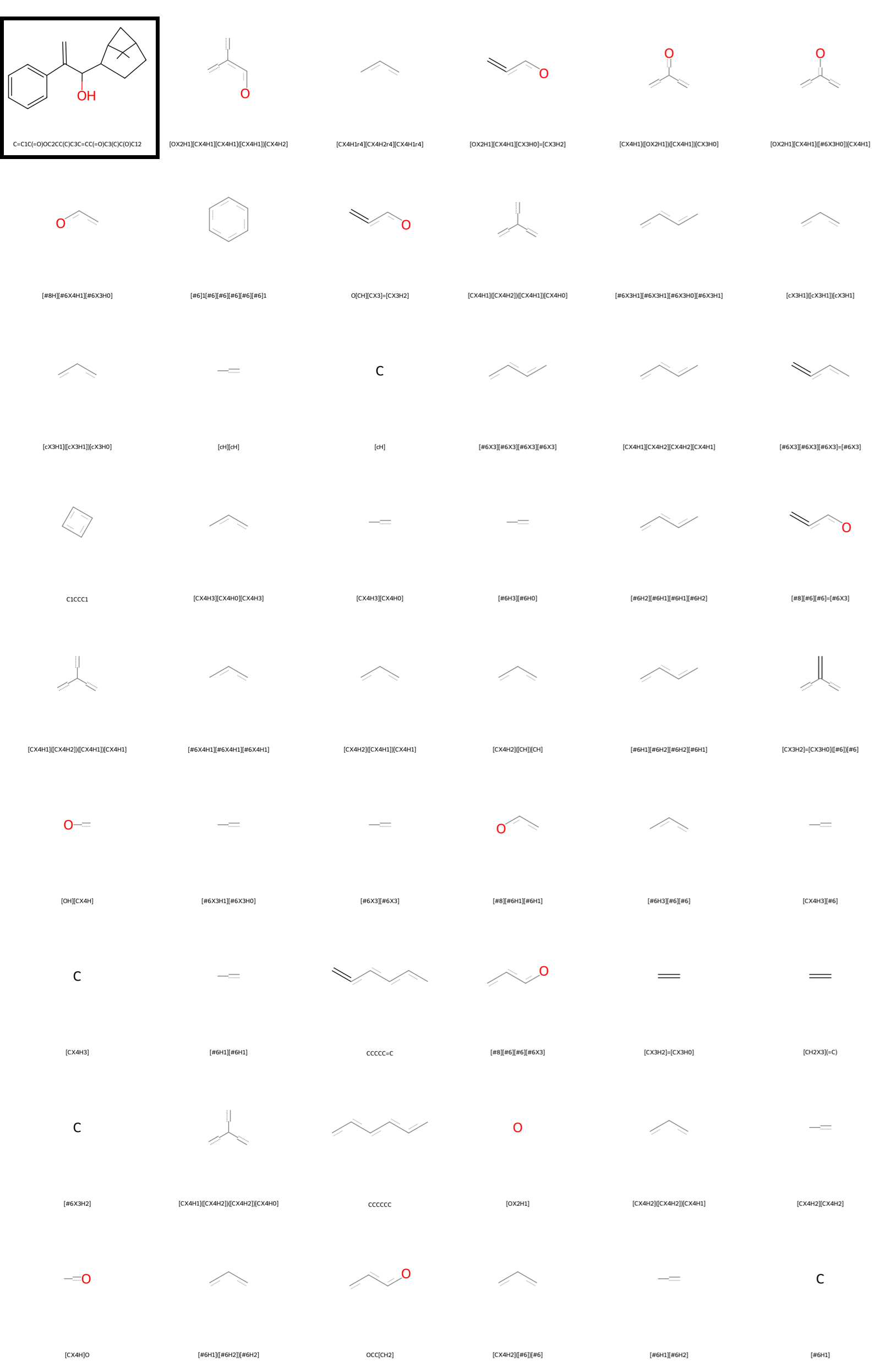}
    \caption{Example of a correctly predicted molecule (top left, black box) from the substructure-to-structure model and the 53 substructures that were provided as input.}
    \label{fig:SI_molecule_and_substructures}
\end{figure}
\restoregeometry

\newpage
\bibliography{references}